\documentclass[12pt]{article}
\usepackage{graphicx}
\begin{document}
\rightline{NKU-05-SF1}
\bigskip
\begin{center}
{\Large\bf Stability and quasi-normal modes  of  charged  black holes in Born-Infeld gravity}

\end{center}
\hspace{0.4cm}
\begin{center}
Sharmanthie Fernando \footnote{fernando@nku.edu} \& Chad Holbrook \footnote{holbrooc@nku.edu}\\
{\small\it Department of Physics \& Geology}\\
{\small\it Northern Kentucky University}\\
{\small\it Highland Heights}\\
{\small\it Kentucky 41099}\\
{\small\it U.S.A.}\\

\end{center}

\begin{center}
{\bf Abstract}
\end{center}

\hspace{0.7cm} 
In this paper we study the stability and  quasi-normal modes of scalar perturbations of black holes. The static charged  black hole considered here is a solution to Born-Infeld electrodynamics coupled to gravity. We conclude that the black hole is stable. We also compare the stability of it with the linear counter-part, Reissner-Nordstrom black hole. The quasi normal modes are computed using the WKB method. The behavior of these modes with the non-linear parameter, temperature, mass of the scalar field and the spherical index are analyzed in detail.\\

\noindent


\section{Introduction}

In this paper we focus on a black hole arising in Einstein-Born-Infeld gravity.  In Maxwell's theory of electrodynamics, the value of the electric field diverges at the origin where the electric charge is located. In 1930, in order to cure these divergences, Born and Infeld develop a electrodynamic theory where the electric field is finite and its total energy is also finite \cite{born}. In Born-Infeld electrodynamics, the electric field of a point charge is given by, 
$E = \frac{ Q}{ \sqrt{ r^4 + \frac{ Q^2}{\beta^2}}}$ where $\beta$ is known as the Born-Infeld parameter. Recently, the Born-Infeld theory has gained attention since its relation to string/M-theory which is a candidate for quantum gravity. Born-Infeld theory arises naturally in open superstrings and D-branes \cite{leigh}. Also, the Born-Infeld action coupled to a dilaton and an axion field is related to the action with an open superstring and abelian gauge field theory \cite{frad} \cite{mat} \cite{beg}.  In addition, the Born-Infeld action arises in the effective action derscribing the dynamics of vector fields on D-branes \cite{tsey1}. Born-Infeld theory and its relation to string theory has been reviewed in \cite{gib1} and \cite{tsey2}.

Black holes in Born-Infeld gravity is quite different from  black hole solutions to Einstein-Maxwell gravity (Reissner-Nordstrom black holes). In the Reissner-Nordstrom black holes, the singularity at $r =0$ is due to the $\frac{Q^2}{r^2}$ term \cite{chandra}. On the other hand, in Born-Infeld black holes, the singularity is dominated by the $\frac{M}{r}$ term. Also, in Born-Infeld black holes, the possibility exists for the mass $M$ for be positive, negative or zero  as described in section(2). Born-Infeld black holes were first constructed by Garcia et.al. \cite{Garcia} and later Demianski presented a solution \cite{demia}  which differs by a constant to the one by Garcia et.al. Geodesics of test particles around a static Born-Infeld black hole was presented by Breton \cite{nora1}. Born-Infeld black holes in isolated horizons were presented  in \cite{nora2}. A Melvin type Universe describing a magnetic field permeating the Universe in Born-Infeld gravity was described by Gibbons and Herdeiro \cite{gib2}. In the same paper, by using duality properties, they obtained Melvin dyonic Universes.

One of the most important characteristics of black holes are its quasi-normal modes which arises during the perturbations of a black hole.  The frequencies of quasi-normal modes depend only on the parameters of the black hole  such as the mass, charge and the angular momentum.  There are many papers written on quasi-normal modes of black holes for fields of perturbations by differing spins. A good review on the subject is given by Kokkotas et.al \cite{kok1}. In this paper, we are interested in studying quasi-normal modes (QNM) of scalar perturbations of a Born-Infeld black hole.

One of the reasons to study QNM's are the results in Loop quantum gravity. It has been observed that the real part of high overtones of QNM's are related to a parameter in Loop quantum gravity called the Barberr-Imirizi parameter. Some of the work done on this subject  are in \cite{hod} \cite{corichi} \cite{mot1} \cite{dreyer} \cite{jo1} \cite{jo2} \cite{mot2} \cite{van}  \cite{set1} \cite{set2} \cite{kun} \cite{ander} \cite{elias}. Also, there have been lot of interest in studying QNM's of AdS black holes due to the AdS/CFT correspondence. Few of the work we would like to mention here related to this are \cite{horo} \cite{car1}  \cite{wang}  \cite{kon1}.

The paper is presented as follows: In section 2, the Born-Infeld black hole solutions are introduced. In section 3, the scalar perturbations are given. In section 4, we will computer the QNM's and discuss the results. Finally, the conclusion is given in section 5.


\section{Static charged  black hole in Einstein-Born-Infeld gravity}

In this section, an introduction to the static charged  black hole
in Einstein-Born-Infeld gravity is given. The most general action for a theory with non-linear electrodynamics coupled to gravity  is as follows:
\begin{equation}
S = \int d^4x \sqrt{-g} \left[ \frac{R }{16 \pi G} + L(F) \right]
\end{equation}
Here, $L(F)$ corresponds to the Lagrangian for the non-linear electrodynamics. For Born-Infeld electrodynamics, $L(F)$ is given by,
\begin{equation}
L(F) = 4 \beta^2 \left( 1 - \sqrt{ 1 + \frac{ F^{\mu \nu}F_{\mu \nu}}{ 2 \beta^2}} \right)
\end{equation}
Here  $F_{\mu \nu}$ is the electromagnetic field strength and $\beta$ is the Born-Infeld parameter. $\beta$ has the dimensions of  $\frac{1}{l^2}$. We will assume $16 \pi G = 1$ in the coming sections. In  the  weak field limit, $L(F)$ has 
the form 
\begin{equation}
L(F) = - F^{\mu \nu}F_{\mu \nu} + O(F^4)
\end{equation}
Note that when $\beta \rightarrow \infty$,  $L(F)$ approaches the one for Maxwell's  electrodynamics given by $ - F^2$.  Out of all theories of non-electrodynamics, Born-Infeld theory has a special place that it is also invariant under electric-magnetic duality which is  discussed in detail by Gibbons and Rasheed \cite{rasheed1}.

For static spherically symmetric case, the electric field $E$ is given by,
\begin{equation}
F_{tr} = E(r) = - \frac{Q}{\sqrt{ r^4 + \frac{Q^2}{\beta^2}}}
\end{equation}
In  this case the  non-linear Lagrangian reduces to,
\begin{equation}
 L(F) = 4 \beta^2 \left( 1 - \sqrt{ 1 - \frac{E^2}{\beta^2}} \right)
\end{equation} 
Due to the square root in the term, $|E|$ has to be smaller than $\beta$. This is the major difference between Born-Infeld electrodynamics and Maxwell electrodynamics.

The static charged black hole with spherical symmetry for the above theory can be obtained as,
\begin{equation}
ds^2 = g(r) dt^2 - g(r)^{-1} dr^2 - r^2 ( d \theta^2 + sin^2(\theta) d \varphi^2)
\end{equation}
with,
\begin{equation}
g(r) = 1 - \frac{2M}{r} + 2 \beta \left( \frac{r^2 \beta}{3} + \frac{1}{r} \int_{r}^{\infty}  \sqrt{ Q^2 + r^4 \beta^2} \right)
\end{equation}
The integral in $g(r)$ can be written in terms of hypergeometric functions $_2F_1$ as
\begin{equation}
g(r) = 1 - \frac{2M}{r} + \frac{2 \beta r^2}{3} \left( 1 - \sqrt{ 1 + \frac{Q^2}{r^4 \beta^2}} \right) + \frac{ 4 Q^2}{ 3 r^2} \hspace{0.2cm}   _2F_1 \left( \frac{1}{4}, \frac{1}{2}, \frac{5}{4}, -\frac{Q^2}{ \beta^2 r^4} \right)
\end{equation}
In the limit $\beta \rightarrow \infty$, $g(r)$  can be expanded to give,
\begin{equation}
g(r) = 1 - \frac{2 M}{r} + \frac{ Q^2}{r^2}
\end{equation}
which is  the function $g(r)$ for the Reissner-Nordstrom black hole for Maxwell's electrodynamics.
Near the origin, the function $g(r)$ can be expanded as,
\begin{equation}
g(r) \approx 1 - \frac{( 2M - \alpha)}{r} - 2 \beta Q + \frac{2 \beta^2}{3} r^2  + \frac{ \beta^3}{5}  r^4
\end{equation}
Here,
\begin{equation}
\alpha = \sqrt{ \frac{\beta}{ \pi} } Q^{3/2} \Gamma \left(\frac{1}{4} \right)^2
\end{equation}
By observing eq.(10), one can see that depending on the values of $M$ and $\alpha$, the sign for the $\frac{1}{r}$ term can be positive, negative or zero. Therefore, on can have time-like or space-like singularity depending on the sign of the $\frac{1}{r}$ term.  For $ 2 M- \alpha =0$, the metric is finite at $r=0$. This is the reason why the Born-Infeld black holes are interesting than the Reissner-Nordstrom black holes.

In the following figure we have sketched the function $g(r)$ for both Born-Infeld and Reissner-Nordstrom black hole. This is a case where Born-Infeld black hole has a single root.

\begin{center}
\scalebox{.9}{\includegraphics{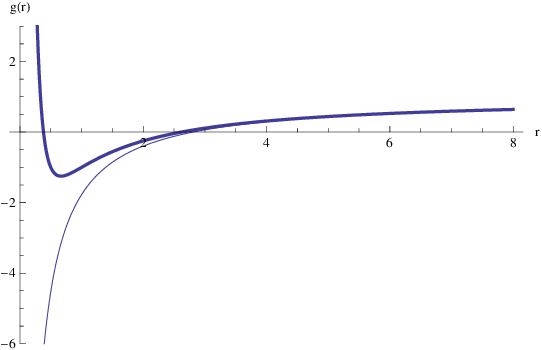}}

\vspace{0.3cm}
\end{center}

Figure 1. The figure shows the function $g(r)$  for $M=1.5$ and $Q=1$. The lighter graph shows the Born-Infeld black hole and the darker one shows the Reissner-Nordstrom black hole.\\

\noindent
A  description of the characteristics of the Born-Infeld black hole is given in \cite{rasheed2} and \cite{nora1}. The Hawking temperature $T_H$ of the black hole is given by,
\begin{equation}
T = \frac{1}{4\pi} \left[ \frac{1}{r_h}  + 2\beta \left( r_h \beta - \frac{\sqrt{(Q^2 + r_h^2 \beta^2)}}{r_h} \right) \right]
\end{equation}
Here, $r_h$ is the event horizon of the black hole. The zeroth and the first law of Born-Infeld black holes  are discussed in detail in \cite{rasheed2}. Static charged black hole solution to Born-Infeld garvity with a cosmological constant was presented in \cite{fer11} \cite{cai} and was extended to higher dimensions in \cite{dey}.


\section{Scalar perturbation of charged Born-Infeld black holes}

In this section, we will develop the equations for a scalar field in the background of the static charged black hole introduced in the previous section. The general equation for a massless scalar field in curved space-time can be written as,
\begin{equation}
\bigtriangledown ^2 \Phi  =0
\end{equation}
which is also equal to,
\begin{equation}
\frac{1}{\sqrt{-g}} \partial_{\mu} ( \sqrt{-g} g^{\mu \nu} \partial_{\nu} \Phi ) =0
\end{equation}
Using the ansatz,
\begin{equation}
\Phi =  e^{- i \omega t} Y(\theta,\phi) \frac{\xi(r)}{r} 
\end{equation}
eq.(14) leads to the radial equation,
\begin{equation}
\left(\frac{d^2}{dr_{*}^2} + \omega^2 \right) \xi(r) = V(r_*) \xi(r)
\end{equation}
where,
\begin{equation}
V(r) =  \frac{l(l+1)}{r^2} g + \frac{g g' }{r} 
\end{equation}
and $r_*$ is the well known ``tortoise'' coordinate given by,
\begin{equation}
dr_{*} = \frac{dr}{g}
\end{equation}
Since $g(r)$ is not in closed from, $r_h$ cannot be evaluated explicitly. 
Note that $l$ is the spherical harmonic index. 
Hence, when $r \rightarrow \infty$, $r_* \rightarrow \infty$ and when $r \rightarrow r_h$, $r_* \rightarrow - \infty$.

The effective potential $V$ for the Born-Infeld black hole is plotted to show how it changes with charge $Q$ and the non-linear parameter $\beta$ in the following figures.

\begin{center}
\scalebox{.9}{\includegraphics{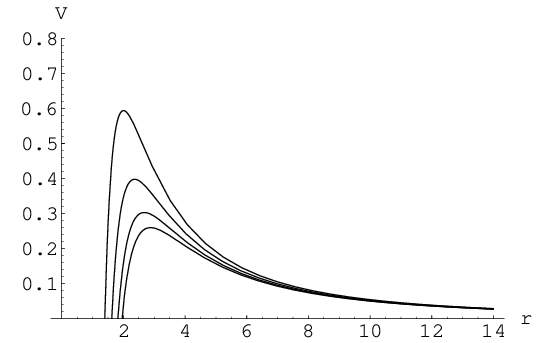}}

\vspace{0.3cm}
\end{center}

Figure 2. The behavior of the effective potential $V(r)$ with 
the charge for the Born-Infeld black hole. Here, $M=1$, $\beta=0.2$ and $l=2$. The height of the potential decreases when the charge decreases.

\begin{center}
\scalebox{.9}{\includegraphics{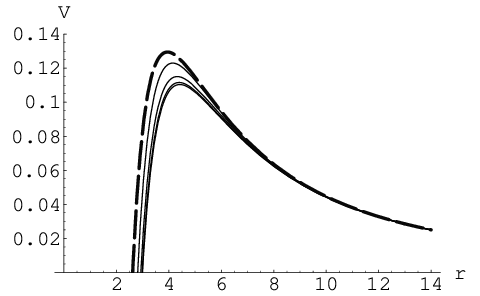}}

\vspace{0.3cm}
\end{center}

Figure 3. The behavior of the effective potential $V(r)$ with 
the non-linear parameter $\beta$. Here, $M=1.5$, $Q=1$ and $l=2$. The maximum height of the potential increases as $\beta$ increases. The dashed one is the potential for the Reissner-Nordstrom black hole with same mass and charge.

\subsection{Remarks on stability}

By observing that the  potentials are real and positive outside the event horizon, it can be concluded that the black holes are stable following the arguments by Chandrasekhar \cite{chandra}. Note that in \cite{fer1}, it was shown to be stable for gravitational perturbations.


\section{Quasi-normal modes of the Born-Infeld black hole}

Quasi-normal modes (QNM) of a classical perturbation of black hole space-times are obtained by studying the solutions to the corresponding wave equation.  In order to obtain the frequencies of QNM, one has to impose boundary conditions on the solutions: for asymptotically flat space-times, the wave is purely ingoing at the horizon and purely out going at the asymptotic infinity.

The wave equations of the black hole perturbations are not easy to solve exactly. Only few cases have been solved exactly: in 2+1 dimensions, the authors can give few examples such as the BTZ black hole \cite{bir1} and the charged dilaton black holes  \cite{fer2} \cite{fer3} \cite{fer4} \cite{fer7}.  Vector perturbations in five dimensions were solved exactly in \cite{nun}.

In order to compute QNM frequencies, we will follow a semi analytical technique developed by Iyer and Will \cite{will}. The method makes use of the WKB approximation. The basics of this method is reviewed  in \cite{fer5}. This approach has been applied to compute QNM's of many black holes. Scalar perturbations of charged dilaton black hole \cite{fer3}, charged scalar perturbations of Reissner-Nordstrom black hole \cite{kon2} and scalar perturbations of acoustic black holes \cite{berti1} are a few of such studies. 

Let $\omega$ be represented as  $\omega = \omega_R - i \omega_I$.  The lowest quasi normal mode $\omega(0)$ of the Born-Infeld black holes are computed. 

First, the quasi normal modes are computed to see the behavior with the non-linear parameter  $\beta$ and graphed in the following figure.

\vspace{0.3cm}
\begin{center}
\scalebox{.9}{\includegraphics{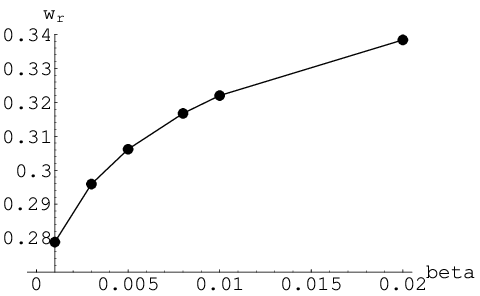}}

\vspace{0.3cm}

\end{center}

Figure 4. The behavior of Re $\omega$ with the non-linear parameter $\beta$ for $M=2$, $Q=1$ and $l=2$.

\begin{center}
\scalebox{.9}{\includegraphics{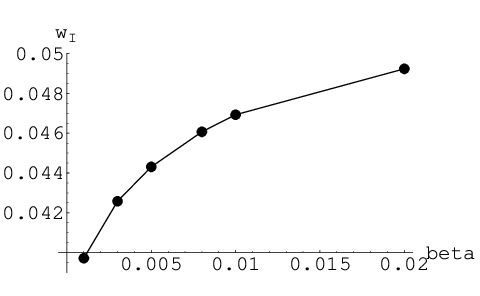}}

\vspace{0.3cm}

\end{center}

Figure 5. The behavior of Im $\omega$ with the non-linear parameter $\beta$ for $M=2$, $Q=1$ and $l=2$
\\

Next, we have computed the QNM's by varying the charge of the black hole. We also computed the QNM's for the Reissner-Nordstrom black hole with the same mass and the charge. Both results are plotted in the following figures.

\begin{center}
\vspace{0.3cm}

\scalebox{.9}{\includegraphics{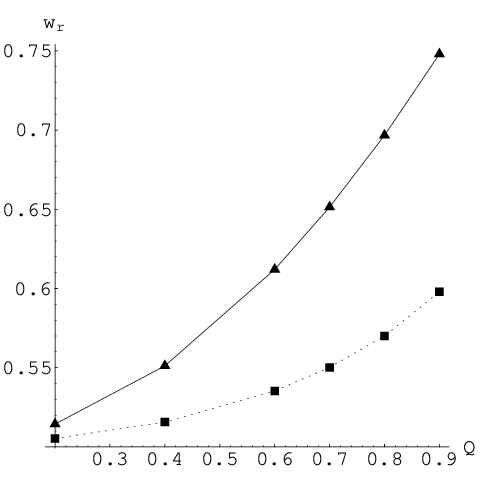}}

\vspace{0.3cm}
\end{center}

Figure 6. The behavior of Re $\omega$ with the charge $Q$ for $M=1$, $\beta=0.04$ and $l=2$. The dotted lines show the graph for Reissner-Nordstrom and the dark lines show the one for the Born-Infeld black hole.

\begin{center}
\scalebox{.9}{\includegraphics{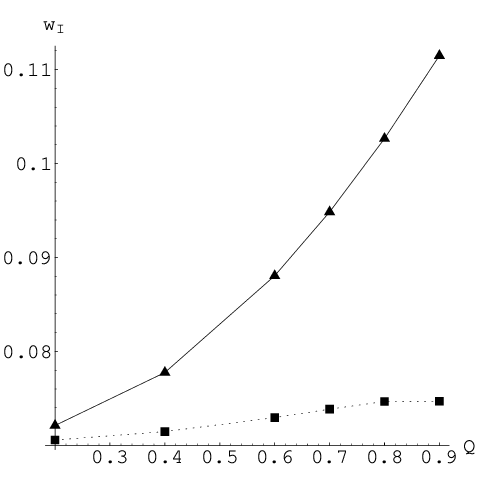}}

\vspace{0.3cm}

\end{center}

Figure 7. The behavior of Im $\omega$ with the charge $Q$ for $M=1$, $\beta=0.04$ and $l=2$. The dotted lines show the graph for Reissner-Nordstrom and the dark lines show the one for the Born-Infeld black hole.\\

For larger values of the  $Q$,   the decay of the scalar field is faster. From the above plots, one can observe that the decay of the scalar field in Born-Infeld black hole is faster than the Reissner-Nordstrom black hole for this particular values of the parameters. One can also conclude that the Born-Infeld black hole are stable since the fields decay in this background. Note that we have done these computations for a Born-Infeld back hole which behaves like the Schwarzschild black hole near the origin. It is necessary to do a evaluation based on all the parameters to fully understand the  behavior of the $\omega_I$ to see how the stability compare with the Reissner-Nordstrom black hole.

We also studied the behavior of the quasi normal modes with spherical index $l$ as given in the following figure.

\vspace{0.3cm}
\begin{center}
\scalebox{.9}{\includegraphics{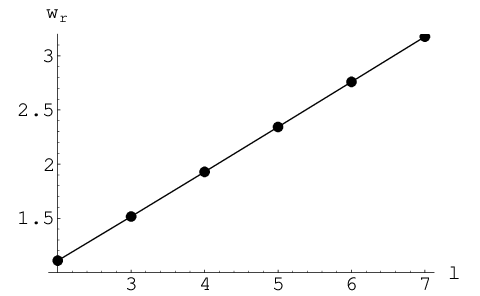}}

\vspace{0.3cm}
\end{center}

Figure 8. The behavior of Re $\omega$ with the  spherical index $l$ for $M=2$, $Q=1$ and $\beta=1$

\begin{center}
\scalebox{.9}{\includegraphics{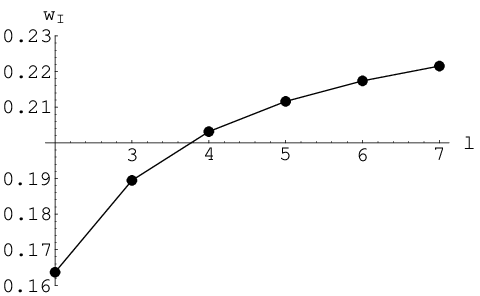}}

\vspace{0.3cm}

\end{center}

Figure 9. The behavior of Im $\omega$ with the spherical index $l$ for $M=2$, $Q=1$ and $\beta=1$\\

It is observed that when the spherical index $l$ is increased, the Re $\omega$ increases and Im $\omega$ approaches a fixed value. This is similar to the behavior observed for charged scalar dilaton black holes in \cite{fer4}.

Next, we have studied the behavior of the Im $\omega$ with temperature of the black hole. The plot of $Im (\omega)$ vs temperature is given in the Figure 9.
One can see a linear behavior of Im $\omega$ with the temperature. This behavior is similar to the Schwarzschild-anti-de-Sitter black hole studied by Horowitz and Hubeny \cite{horo}.

\begin{center}
\vspace{0.3 cm}

\scalebox{.9}{\includegraphics{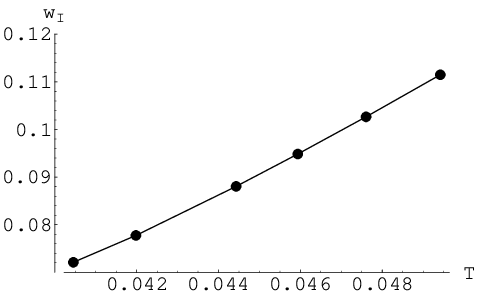}}

\vspace{0.3cm}
\end{center}

 Figure 10. The behavior of $\omega_I$ with the Temperature for  $l=2$. The mass $M =1$ and $\beta = 0.04$.


\subsection{Asymptotic  quasi normal modes}

One of the reasons to attract attention to QNM's  computation in the recent past has to do with the conjecture that the real part  of QNM's corresponds to the minimum energy change of a quantum transitions of a black hole \cite{hod} \cite{dreyer}. There have been a number of works done to compute the high frequency QNM's and the asymptotic value of the real part of QNM frequencies. Recently, Das and Shankaranarayanan developed a method to compute QNM frequencies of $(D+2)$ dimensional spherically symmetric single horizon black holes with generic singularities \cite{das}. In this paper, we apply this method to compute the asymptotic values of the QNM's of the Born-Infeld black hole. First, we will summarize their derivation as follows: Consider a general $(D+2)$ dimensional spherically symmetric line element given by,
\begin{equation}
ds^2= \gamma(r) dt^2 + \frac{dr^2}{\eta(r)} + r^2 d \Omega^D
\end{equation}
Suppose there is a singularity of the line element at $r \rightarrow 0$, then near  the singularity the function $\gamma(r)$ and $\eta(r)$ have the power law behavior as,
$$ \gamma(r) \approx r^{\frac{ 2p }{q}}$$
\begin{equation}
\frac{1}{\eta(r)} \approx  r^{ \frac{2 (p-q+2)}{q}}
\end{equation}
This was shown in the works of Szekeres-Iyer \cite{sze1} \cite{sze2}.
For scalar perturbations, Das and Shankaranarayanan obtained,
\begin{equation}
\omega_{QNM} = \frac{i} { 2 c_0} ( n + \frac{1}{2} ) \pm \frac{ 1}{ 4 \pi c_0} Log \left( 1 + 2 Cos( \frac{\pi}{2}(Dq-2) ) \right)
\end{equation}
Here, $c_0$ is a constant and is related to the temperature. For the Born-Infeld black hole $c_0 = \frac{1}{4 \pi T}$.

Now, we will apply this method to compute the QNM's of the Born-Infeld black hole. Recall that closer to the singularity $ r \rightarrow 0$, the function $\gamma(r)$ is given in eq.(10). Hence a power law behavior for $\gamma(r)$ and $\eta(r)$ can be obtained as,
\begin{equation}
\gamma(r) \approx a_1 r^{-1}  \Rightarrow  \frac{ 2 p }{q} = -1
\end{equation}
Also,
\begin{equation}
\frac{1}{\eta(r)} = \frac{1}{\gamma(r)} \approx \frac{r}{a_1} \Rightarrow 2(\frac{p}{q} - 1 +\frac{2}{q}) =1
\end{equation}
Note that $a_1 = -(2 M - \alpha)$. Now from eq.(22) and eq.(23), $p$ and $q$ values  for the Born-Infeld black hole can be computed to be,
\begin{equation}
p = -\frac{1}{2}; \hspace{1 cm} q = 1
\end{equation}
From eq.(21), the QNM's  are  given by,
\begin{equation}
\omega_{QNM} =  i  2 \pi T_H ( n + \frac{1}{2} ) \pm  T_H Log (3)
\end{equation}
The real part of $\omega$ behaves as $T_H Log (3)$ according to this derivation. In fact in \cite{das} it was derived that $D+2$ dimensional Schwarzschild, and stringy black holes in 4 and 5 dimensions also have the same behavior for real $\omega$.


\subsection{Massive scalar perturbations}

It has been observed that the massive modes decay slower than the massless field for the Schwarzschild black hole in \cite{lior} \cite{kon4}. An interesting question was posed whether the decreasing decay rate can approach zero leading to a long living mode. Here we are computing the QNM's  for a massive scalar field in the Born-Infeld black hole to see if this behavior persist in this case.

The general equation for a massive scalar field in curved space-time is written as,
\begin{equation}
\bigtriangledown ^2 \Phi - m^2 \Phi =0
\end{equation}
Using the ansatz,
\begin{equation}
\Phi =  e^{- i \omega t} Y(\theta,\phi) \frac{\xi(r)}{r} 
\end{equation}
eq.(26) leads to the radial equation given in eq.(16) with the modified potential,
\begin{equation}
V(r) =  \frac{l(l+1)}{r^2} g + \frac{g g' }{r} + m^2 g(r)
\end{equation}
The effective potential $V$ is plotted by varying the mass $m$ of the scalar field in the following figure.

\begin{center}
\vspace{0.3 cm}

\scalebox{.9}{\includegraphics{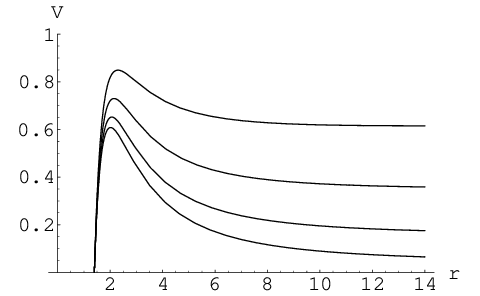}}

\vspace{0.3cm}
\end{center}

Figure 11. The behavior of $V$ with the mass $m$ of the scalar field. Here $M=1$, $Q=0.8$, $\beta=0.2$ and $l=2$. When the mass decreases, the height of the potential also decreases.\\

We computed the QNM's for the massive scalar field decay using the WKB approximation discussed in section 4. They are given in the following figures.
\vspace{0.3cm}
\begin{center}
\scalebox{.9}{\includegraphics{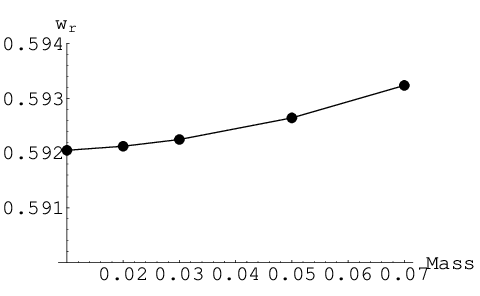}}

\vspace{0.3cm}
\end{center}

Figure 12. The behavior of Re $\omega$ with the  mass of the scalar field $m$ for $M=1$, $Q=1$, $\beta=0.04$ and $l=2$ 

\begin{center}
\scalebox{.9}{\includegraphics{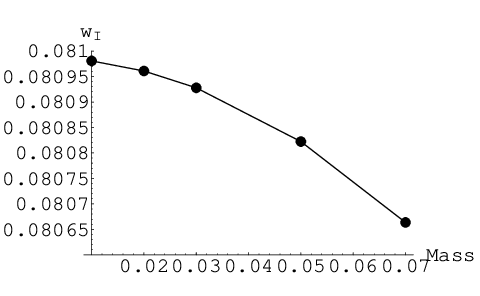}}

\vspace{0.3cm}

\end{center}

Figure 13. The behavior of Im $\omega$ with the mass of the scalar field $m$ for  $M=1$, $Q=1$, $\beta=0.04$ and $l=2$.\\

By observing the behavior of Im $\omega$ when the mass $m$ increases it is clear that the decay is slower for massive scalar field similar to the Schwarzschild case. On the other hand, the frequency of oscillations of the modes seems to increase with the mass.

\section{Conclusion}
We have studied quasi normal modes for the charged  black holes in Einstein-Born-Infeld gravity  with scalar perturbations. The lowest quasi normal modes are computed  using a WKB method.  

The main result of this paper is that Born-Infeld black holes are stable under scalar perturbations. It is also noted that the scalar field decay faster in the Born-Infeld black hole background in comparison with the Reissner-Nordstrom black hole. We have done this study only for a particular class of parameters and a detailed analysis is necessary to predict the behavior for all parameters.

We obtained an approximation to the asymptotic QNM's  using a method developed in \cite{das}. There, the linear behavior of $\omega$ with the temperature was observed. The computed $\omega_I$ using the WKB method support this. Similar behavior  was observed first by Horowitz and Hubeny for Schwarzschild-anti-de-Sitter black holes \cite{horo}. There, they showed a linear relation between QNM's and temperature for large black holes in several dimensions. For black holes, in anti-de-Sitter space, relations between QNM's and conformal field theory of the boundary are discussed in many papers. 

It  is observed that the spherical index $l$ is increased, the Re $\omega$ increases leading to greater oscillations.  However, Im $\omega$ approaches  a fixed value for larger $l$. This is similar to what was observed in the static charged dilaton black hole in \cite{fer4}.

Even though the paper's focus was on massless scalar field, we also analyzed the behavior of a massive scalar field. It decays slower than the massless field, similar to the observation of other black holes \cite{kon4}.

A natural extension of this work is to study the stability of the Born-Infeld black hole with a cosmological constant \cite{fer1}. In particular, the Born-Infeld-Ads case is worthy of study due to the AdS/CFT conjecture. The electrically charged black hole in anti-de Sitter space has been shown to be unstable for large black holes by using linear perturbation techniques \cite{gub}. It would be interesting to study how the non-linear nature effects the instability of such solutions.

The supersymmetric nature of the Born-Infeld black hole would be  interesting to investigate. The extreme Reissner-Nordstrom  black hole can be embedded in $N=2$ supergravity theory \cite{gib3} \cite{gib4}. Onozawa  et.al. \cite{ono} showed that the QNM's of the extreme RN black hole for spin 1, 3/2 and 2 are the same. A suitable supergravity theory to embed the Born-Infeld black hole has not been constructed. However, one can do a similar study such as in Onozawa et. al. \cite {ono} in terms of the QNM's for extreme Born-Infeld black holes. From the eq.(17)  when $ 2 M - \alpha$ is smaller than zero, the black holes behave similar to the Reissner-Nordstrom black hole. Hence, it is possible to construct extreme case for Born-Infeld black hole by choosing appropriate parameters. 

The WKB approach considered here was extended to the sixth order by Konoplya \cite{kon6} which gives greater accuracy in computing the QNM frequencies. It would be interesting to use it to compute frequencies of  high overtones.

\vspace{0.5cm}

{\bf Acknowledgments}: This work was supported in part by a grant from the Kentucky Space Grant Consortium.


\end{document}